\documentclass[12pt]{article}
\usepackage[a4paper]{geometry}
\usepackage[english]{babel}

\usepackage[pagebackref,draft=false]{hyperref}
\hypersetup{colorlinks,
	linkcolor=myrefcolor,
	citecolor=mycitecolor,
	urlcolor=myurlcolor}

\usepackage{caption}
\usepackage{etaremune}

\usepackage{xcolor}
\definecolor{myurlcolor}{rgb}{0,0,0.4}
\definecolor{mycitecolor}{rgb}{0,0.5,0}
\definecolor{myrefcolor}{rgb}{0.5,0,0}
\usepackage{graphicx}
\usepackage{tikz}
\usepackage{tikz-cd}
\usepackage{mathrsfs}

\usepackage{etoolbox}
\usepackage{makeidx}
\usepackage{sectsty}
\usepackage{dsfont}
\usepackage{enumitem} 
\usepackage[]{latexsym}
\usepackage{braket}
\usepackage{caption}
\usepackage[utf8]{inputenx}
\usepackage[T1]{fontenc}
\usepackage{lmodern}
\usepackage{textcomp}
\usepackage{microtype}
\usepackage{totcount}
\usepackage{blindtext}

\usepackage{fancyhdr}
\usepackage{amsmath}
\usepackage{amsfonts}
\usepackage{amstext}
\usepackage{amssymb}
\usepackage{amscd}

\usepackage{xcolor}
\usepackage{mathrsfs}
\usepackage{tikz}
\usepackage{tikz-cd}

\usepackage{fancyhdr}

\newcommand{\be}{\begin{equation}}
\newcommand{\ee}{\end{equation}}

\newcommand{\bea}{\begin{eqnarray}}
\newcommand{\eea}{\end{eqnarray}}

\newcommand{\blue}[1]{\textcolor{blue}{{#1}}}

\newcommand{\hh}{\mathcal{H}}
\newcommand{\bh}{\mathcal{B}(\mathcal{H})}

\newcommand{\Uh}{\mathcal{U}(\mathcal{H})}

\newcommand{\stsph}{\mathscr{S}(\mathcal{H})}
\newcommand{\Tr}{\textit{Tr}}

\newtheorem{remark}{Remark}

\begin{document}

	\title{Unfolding of relative $g$-entropies and monotone metrics}

	\author{F. Di Nocera$^{ 1, 2}$\href{https://orcid.org/0000-0002-1415-2422}{\includegraphics[scale=0.7]{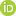}}, \\
		\footnotesize{$^{1}$\textit{ Max Planck Institute for Mathematics in the Sciences, Leipzig, Germany}} \\
		\footnotesize{$^{2}$\textit{ e-mail: \texttt{fabiodncr[at]gmail.com} and \texttt{dinocer[at]mis.mpg.de}}} \\ 
	}
	
	\maketitle

\begin{abstract}
	We discuss the geometric aspects of a recently described unfolding procedure and show the form of objects relevant in the field of Quantum Information Geometry in the unfolding space. In particular, we show the form of the quantum monotone metric tensors characterized by Petz and retrace in this unfolded perspective a recently introduced procedure of extracting a covariant tensor from a relative $g$-entropy.
\end{abstract}
%%%%%%%%%%%%%%%%%%%%%%%%%%%%%%%%%%%%%%%%%%%
%%%%%%%%%%%%%%%%%%%%%%%%%%%%%%%%%%%%%%%%%%
%%%%%%%%%%%%%%%%%%%%%%%%%%%%%%%%%%%%%%%%%%
%%%%%%%%%%%%%%%%%%%%%%%%%%%%%%%%%%%%%%%%%%
%%%%%%%%%%%%%%%%%%%%%%%%%%%%%%%%%%%%%%%%%%
%%%%%%%%%%%%%%%%%%%%%%%%%%%%%%%%%%%%%%%%%%
%%%%%%%%%%%%%%%%%%%%%%%%%%%%%%%%%%%%%%%%%%
%%%%%%%%%%%%%%%%%%%%%%%%%%%%%%%%%%%%%%%%%%
%%%%%%%%%%%%%%%%%%%%%%%%%%%%%%%%%%%%%%%%%%
%%%%%%%%%%%%%%%%%%%%%%%%%%%%%%%%%%%%%%%%%%
\tableofcontents
%%%%%%%%%%%%%%%%%%%%%%%%%%%%%%%%%%%%%%%%%%
%%%%%%%%%%%%%%%%%%%%%%%%%%%%%%%%%%%%%%%%%%
%%%%%%%%%%%%%%%%%%%%%%%%%%%%%%%%%%%%%%%%%%
%%%%%%%%%%%%%%%%%%%%%%%%%%%%%%%%%%%%%%%%%%
%%%%%%%%%%%%%%%%%%%%%%%%%%%%%%%%%%%%%%%%%%
%%%%%%%%%%%%%%%%%%%%%%%%%%%%%%%%%%%%%%%%%%
%%%%%%%%%%%%%%%%%%%%%%%%%%%%%%%%%%%%%%%%%%
%%%%%%%%%%%%%%%%%%%%%%%%%%%%%%%%%%%%%%%%%%
%%%%%%%%%%%%%%%%%%%%%%%%%%%%%%%%%%%%%%%%%%
%%%%%%%%%%%%%%%%%%%%%%%%%%%%%%%%%%%%%%%%%%
%%%%%%%%%%%%%%%%%%%%%%%%%%%%%%%%%%%%%%%%%%
\section{Introduction}
%%%%%%%%%%%%%%%%%%%%%%%%%%%%%%%%%%%%%%%%%%
%%%%%%%%%%%%%%%%%%%%%%%%%%%%%%%%%%%%%%%%%%
Fisher-Rao metric tensor \cite{Rao-1945} on classical statistical models can be seen as a ``second-order expansion'' of Kullback-Leibler relative entropy between points in the model. A more general fact actually holds, namely that it is possible to extract covariant tensors from second-order derivatives of \emph{relative entropies} \cite{Amari-2016} satisfying the monotonicity property  with respect to Markov morphisms \cite{Csizar-1963}. This is in perfect agreement with a crucial result obtained by Cencov \cite{Cencov-1982}, i.e. the characterization of the Fisher-Rao metric tensor as the only metric tensor which is invariant under \emph{congruent embeddings}.

In the quantum case, however, the situation is more complex. In \cite{Petz-1996}, (developing on the work of Cencov and Morozova in \cite{C-M-1991}) Petz showed that there is a family of quantum generalizations of Fisher-Rao metric tensor labelled by operator monotone functions \cite{Loewner-1934}. This characterization is largely used in the context of Quantum Information Geometry and was obtained considering monotonicity under the class of \emph{completely positive, trace preserving} (CPTP) \emph{maps}, that are the appropriate generalization to the quantum case of Markov's maps \cite{Choi-1975}.

In \cite{L-R-1999}, the authors show that the characterization obtained by Petz can be recovered by considering some ``second-order expansion'' of a family of relative entropies labelled by a convex operator function \cite{Kraus-1936} $g$ defined on density matrices, called \emph{relative g-entropies}.

Our main goal is to retrace these basic facts about Quantum Information Geometry in an unfolded framework described in \cite{C-DC-DN-V-2022, M-M-V-V-2017, C-DC-L-M-M-V-V-2018}. The rough idea behind the unfolding procedure is to specify a density operator $\rho$ by means of its spectrum and a unitary operator $U$ that diagonalizes $\rho$. Then since density matrices are positive semi-definite, trace-one matrices, their spectrum is specified (up to permutations) by a probability distribution over a finite sample space. In the language of Information Geometry, by points on a simplex. This idea will be made clear in section \ref{sec:unfolding}, while in section \ref{sec:unfolded IG} we will show the aforementioned results from \cite{Petz-1996} and \cite{L-R-1999} in this framework.
%%%%%%%%%%%%%%%%%%%%%%%%%%%%%%%%%%%%%%%%%%
%%%%%%%%%%%%%%%%%%%%%%%%%%%%%%%%%%%%%%%%%%
%%%%%%%%%%%%%%%%%%%%%%%%%%%%%%%%%%%%%%%%%%
%%%%%%%%%%%%%%%%%%%%%%%%%%%%%%%%%%%%%%%%%%
%%%%%%%%%%%%%%%%%%%%%%%%%%%%%%%%%%%%%%%%%%
\section{Unfolding of the space of quantum states} \label{sec:unfolding}
%%%%%%%%%%%%%%%%%%%%%%%%%%%%%%%%%%%%%%%%%%
%%%%%%%%%%%%%%%%%%%%%%%%%%%%%%%%%%%%%%%%%%
Let $\hh$ be a finite-dimensional Hilbert space of complex dimension $n$, let also $\bh$ be the space of (bounded) operators acting on the Hilbert space $\hh$, the (real) subspace of $\bh$ of self-adjoint operators will be denoted as $\mathcal{B}_{sa}(\mathcal{H})$. A \emph{mixed quantum state} is defined as a positive semi-definite (hence self-adjoint) operator with trace one. If a mixed state is also invertible, i.e. is positive definite, it is said to be \emph{faithful}. The space of all such operators is called \emph{space of faithful quantum states},
%%%%%%%%%%%%%%%%%%%%%%%%%%%%%%%%%%%%%%%%%%
\begin{equation}
\stsph = \{ \, \rho \in \mathcal{B}_{sa}(\mathcal{H}) \, | \quad  \rho > 0, \quad \Tr{ \, \rho} = 1   \}.
\end{equation}
%%%%%%%%%%%%%%%%%%%%%%%%%%%%%%%%%%%%%%%%%%

For space constraint, here we will not discuss in detail the geometry of the space of quantum states, and refer the reader to \cite{G-K-M-2006, A-S-1999, B-Z-2006}. The space of faithful quantum states is a smooth manifold and our discussion will be restricted to this setting because the monotone quantum metric tensors classified by Petz are well-defined only on this manifold. However, it is worth mentioning that the whole space of quantum states $\overline{\stsph}$ (i.e., the topological closure of $\stsph$ in $\mathcal{B}_{sa}(\hh)$) is a stratified manifold, and that its extremal points (i.e., the pure states) form a smooth manifold diffeomorphic to $\mathbb{CP}( \hh)$ on which the Riemannian structure is fiexd to be the Fubini-Study metric tensor because of the requirement of unitary invariance \cite{DA-F-2021}.

For self-adjoint operators holds a crucial result, known as spectral theorem, that states that for any self-adjoint operator $A \in \mathcal{B}_{sa} ( \hh)$ there exist a orthonormal basis of $\hh$ consisting of eigenvectors of $A$. This means that we can always find a basis in which the self-adjoint operator is diagonal. Moreover, the eigenvalues are real numbers. 

Change of basis on an Hilbert space $\hh$ are performed by means of an action of unitary operators on $\hh$. An operator $U \in \bh$ is said to be \emph{unitary} if $U U^\dagger = U^\dagger U = \mathbb{I}$. For these reasons, we can write any mixed state $\rho$ with eigenvalues $\{ \lambda_1, \lambda_2,\dots,\lambda_n\}$ as
%%%%%%%%%%%%%%%%%%%%%%%%%%%%%%%%%%%%%%%%%%	
\begin{equation}
\rho = U \rho_0 U^\dagger,
\end{equation}
%%%%%%%%%%%%%%%%%%%%%%%%%%%%%%%%%%%%%%%%%%
with $\rho_0$ given, with respect to some reference basis in $\hh$, by
%%%%%%%%%%%%%%%%%%%%%%%%%%%%%%%%%%%%%%%%%%
\begin{equation}
\rho_0 =
\begin{bmatrix}
\lambda_1 & 0 & \dots & 0 \\
0 & \lambda_2 & \dots & 0 \\
\vdots & \vdots & \ddots & \vdots \\
0 & 0 & \dots & \lambda_n
\end{bmatrix}.
\end{equation}
%%%%%%%%%%%%%%%%%%%%%%%%%%%%%%%%%%%%%%%%%%
Notice that the positive definiteness of $\rho_0$ implies that all $\lambda_j$'s are strictly positive, while the condition on the trace implies that their sum is 1. This means that, roughly speaking, the eigenvalues of $\rho$ are the components of a probability distribution on a sample space of cardinality $n$. We now want to start from this observation to introduce our unfolding procedure, in order to do so, let us make this analogy between probability vectors and diagonal states more formal.

The set of probability distributions on a finite sample space of cardinality $n$ can be geometrically represented as the $n-1$-dimensional standard simplex $\Delta_{n-1}$. This can be realized as a subset of $\mathbb{R}^n$ as
%%%%%%%%%%%%%%%%%%%%%%%%%%%%%%%%%%%%%%%%%%
\begin{equation}
\Delta_{n-1} = \{ \mathbf{p} = ( p_1, p_2, \dots, p_n) \in \mathbb{R} \, | \quad \sum_j p_j = 1, \quad p_j \ge 0 \quad \forall j = 1,2,\dots,n \}.
\end{equation}
%%%%%%%%%%%%%%%%%%%%%%%%%%%%%%%%%%%%%%%%%%
We now want to look at probability distributions as diagonal mixed states, basically identifying the components of the vector $\mathbf{p}$ with the eigenvalues of some mixed state. Since we want to restrict our discussion to faithful states, we need to consider only the cases in which the components of $\mathbf{p}$ are strictly positive, meaning that we have to consider the space
%%%%%%%%%%%%%%%%%%%%%%%%%%%%%%%%%%%%%%%%%%
\begin{equation}
\Delta^{\mathrm{o}}_{n-1} = \{ \mathbf{p} = ( p_1, p_2, \dots, p_n) \in \mathbb{R} \, | \quad \sum_j p_j = 1, \quad p_j > 0 \quad \forall j = 1,2,\dots,n \}.
\end{equation}
%%%%%%%%%%%%%%%%%%%%%%%%%%%%%%%%%%%%%%%%%%
In this way, we can immerse $\Delta^{\mathrm{o}}_{n-1}$ in the space of faithful quantum states as follows,
%%%%%%%%%%%%%%%%%%%%%%%%%%%%%%%%%%%%%%%%%%
\begin{equation}
i: \Delta^{\mathrm{o}}_{n-1} \ni \mathbf{p} = ( p_1, p_2,\dots,p_n) \mapsto i (\mathbf{p}) =
\begin{bmatrix}
p_1 & 0 & \dots & 0 \\
0 & p_2 & \dots & 0 \\
\vdots & \vdots & \ddots & \vdots \\
0 & 0 & \dots & p_n
\end{bmatrix}
\in \stsph.
\end{equation}
%%%%%%%%%%%%%%%%%%%%%%%%%%%%%%%%%%%%%%%%%%
%%%%%%%%%%%%%%%%%%%%%%%%%%%%%%%%%%%%%%%%%%
\begin{remark}
	Let us stress here that in the definition of $i$ there is an implicit choice of a basis in $\bh$. In fact, choosing to immerse the probability distribution \emph{diagonally} amounts to choosing a basis $\{ \ket{e_1}, \ket{e_2}, \dots, \ket{e_n} \}$ of $\hh$ and write
	%%%%%%%%%%%%%%%%%%%%%%%%%%%%%%%%%%%%%%%%%%
	\begin{equation}
	i (\mathbf{p}) = \sum_{j=1}^n p^j \ket{e_j} \bra{e_j}.
	\end{equation}
	%%%%%%%%%%%%%%%%%%%%%%%%%%%%%%%%%%%%%%%%%%
	Clearly this implicit choice does not lead to any loss of generality of the discussion.
\end{remark}
%%%%%%%%%%%%%%%%%%%%%%%%%%%%%%%%%%%%%%%%%%
Any diagonal faithful state can be realized as the image via $i$ of some probability distribution, implying that for any faithful state $\rho$ there exists a $\mathbf{p} \in \Delta^{\mathrm{o}}_{n-1}$ and a unitary operator $U$ such that
%%%%%%%%%%%%%%%%%%%%%%%%%%%%%%%%%%%%%%%%%%
\begin{equation}\label{eqn:projection_map_1}
\rho = U \, i(\mathbf{p}) U^\dagger,
\end{equation}
%%%%%%%%%%%%%%%%%%%%%%%%%%%%%%%%%%%%%%%%%%
but it is easy to see that this doesn't specify uniquely neither $\mathbf{p}$ nor $U$. In fact, the action of the unitary group considered has a non.trivial isotropy group, so that $U$ is not uniquely determined. On the other hand $\mathbf{p}$ is determined only up to permutations, the redundancy on $\mathbf{p}$ can be removed by considering equivalence classes of probability distributions with respect to permutations, meaning going to the so called \emph{Weyl chamber}. 
%%%%%%%%%%%%%%%%%%%%%%%%%%%%%%%%%%%%%%%%%%
\begin{remark}
	Let us notice here that while the degeneracy on the choice of the probability distribution can be solved by taking the quotient with respect to the permutation group, the same can not be done for $\Uh$. In fact, it would be possible to quotient with respect to the isotropy group of the considered action if it did not depend on the point $i(\mathbf{p})$ on which it acts, which is not the case for our action.
\end{remark}
%%%%%%%%%%%%%%%%%%%%%%%%%%%%%%%%%%%%%%%%%%
Equation \eqref{eqn:projection_map_1} gives a many-to-one correspondence between elements in the cartesian product $\Delta^{\mathrm{o}}_{n-1} \times \Uh$, i.e. couples $(U, \mathbf{p})$, and faithful states. This is exactly what we want to achieve with our unfolding procedure, $\mathcal{M} (\hh) := \Delta^{\mathrm{o}}_{n-1} \times \Uh$ will be our unfolding space, and we use equation \eqref{eqn:projection_map_1} to define the map
%%%%%%%%%%%%%%%%%%%%%%%%%%%%%%%%%%%%%%%%%%
\begin{equation}\label{eqn:projection_map_2}	
\pi: \Uh \times \Delta^{\mathrm{o}}_{n-1} \ni ( U, \mathbf{p}) \mapsto \pi( U, \mathbf{p}) = U \, i(\mathbf{p}) U^\dagger \in \stsph.
\end{equation}
%%%%%%%%%%%%%%%%%%%%%%%%%%%%%%%%%%%%%%%%%%
The map $\pi$ can be seen to be a smooth projection of $\mathcal{M(H)}$ on the space of faithful states \cite{C-DC-L-M-M-V-V-2018}.

Like for all product spaces, it is possible to project $\mathcal{M}$ on its two factors, in particular, let us consider the projection on the second factor,
%%%%%%%%%%%%%%%%%%%%%%%%%%%%%%%%%%%%%%%%%%
\begin{equation} \label{eqn:definition_dequantization_map}
\pi_D : \Uh \times \Delta^{\mathrm{o}}_{n-1} \ni ( U, \mathbf{p}) \mapsto \mathbf{p} \in \Delta^{\mathrm{o}}_{n-1}.
\end{equation}
%%%%%%%%%%%%%%%%%%%%%%%%%%%%%%%%%%%%%%%%%%
The projection $\pi_D$ will be useful for recognizing classical structures in the unfolded space, in fact, it can be regarded as some sort of ``dequantization map'', taking a mixed quantum state and giving as its image a classical mixture with weights given by the eigenvalues of the mixed state; basically forgetting about the non-classical features of the system.

Since we want to discuss Information Geometry on the unfolding space, a brief discussion regarding the tangent spaces to the unfolding space $\mathcal{M} ( \hh)$ is unavoidable. Since $\mathcal{M} ( \hh)$ is the product of two manifolds, we can consider vectors in $T_{( U, \mathbf{p})}\mathcal{M} ( \hh)$ as couples of vectors tangent to $\Uh$ and $\Delta^{\mathrm{o}}_{n-1}$ respectively \cite{A-M-R-2012}. Vectors tangent to $\Uh$ are given by skew-adjoint matrices \cite{Hall-2013}, while it is easy to see that the tangent space of $\Delta^{\mathrm{o}}_{n-1}$ is isomorphic to $\mathbb{R}^{n-1}$ and given by vectors in $\mathbb{R}^n$ whose components sum to zero.

This means that every $V_{( U, \mathbf{p})} \in T_{ ( U, \mathbf{p})} \mathcal{M} ( \hh)$ can be written as
%%%%%%%%%%%%%%%%%%%%%%%%%%%%%%%%%%%%%%%%%%
\begin{equation}
V_{( U, \mathbf{p})} = ( i H, v)
\end{equation}
%%%%%%%%%%%%%%%%%%%%%%%%%%%%%%%%%%%%%%%%%%
with $H$ a self-adjoint matrix and $v \in \mathbb{R}^n$ such that $\sum_j v^j = 0$.
%%%%%%%%%%%%%%%%%%%%%%%%%%%%%%%%%%%%%%%%%%
%%%%%%%%%%%%%%%%%%%%%%%%%%%%%%%%%%%%%%%%%%
%%%%%%%%%%%%%%%%%%%%%%%%%%%%%%%%%%%%%%%%%%
%%%%%%%%%%%%%%%%%%%%%%%%%%%%%%%%%%%%%%%%%%
%%%%%%%%%%%%%%%%%%%%%%%%%%%%%%%%%%%%%%%%%%
\section{Monotone metrics and relative $g$-entropies on the unfolding space} \label{sec:unfolded IG}
%%%%%%%%%%%%%%%%%%%%%%%%%%%%%%%%%%%%%%%%%%
%%%%%%%%%%%%%%%%%%%%%%%%%%%%%%%%%%%%%%%%%%
%%%%%%%%%%%%%%%%%%%%%%%%%%%%%%%%%%%%%%%%%%
%%%%%%%%%%%%%%%%%%%%%%%%%%%%%%%%%%%%%%%%%%
%%%%%%%%%%%%%%%%%%%%%%%%%%%%%%%%%%%%%%%%%%
Petz's characterization of monotone quantum metrics is used in the context of Quantum Information Theory when considering the possible quantum analogues of Fisher-Rao metric tensor. In \cite{Petz-1996}, Petz characterized all metric tensors on the space of quantum states obeying monotonicity under \emph{completely positive, trace-preserving maps} \cite{Petz-1996} as those metric whose action on the vectors $A,B \in T_\rho \stsph$ can be written as
%%%%%%%%%%%%%%%%%%%%%%%%%%%%%%%%%%%%%%%%%%
\begin{equation} \label{eqn:Petz_characterization}
G^f|_\rho  ( A, B) = \Tr \left( A \left( \mathbf{K}^f_\rho \right)^{-1} ( B )\right).
\end{equation}
%%%%%%%%%%%%%%%%%%%%%%%%%%%%%%%%%%%%%%%%%%
Here $\mathbf{K}^f_\rho$ is a superoperator depending on the point $\rho$ and on the operator monotone function $f$. The superoperator $\mathbf{K}_f$ is defined as
%%%%%%%%%%%%%%%%%%%%%%%%%%%%%%%%%%%%%%%%%%
\begin{equation}
\mathbf{K}^f_\rho = f(\mathbf{L}_\rho\mathbf{R}_{\rho^{-1}}) \mathbf{R}_{\rho},
\end{equation}
%%%%%%%%%%%%%%%%%%%%%%%%%%%%%%%%%%%%%%%%%%
where $\mathbf{L}_\rho$ and $\mathbf{R}_{\rho}$ are respectively the left and right multiplications by means of $\rho$, namely
%%%%%%%%%%%%%%%%%%%%%%%%%%%%%%%%%%%%%%%%%%
\begin{equation}
\mathbf{L}_\rho ( B) = \rho B, \quad \mathbf{R}_{\rho} ( B) = B \rho.
\end{equation}
%%%%%%%%%%%%%%%%%%%%%%%%%%%%%%%%%%%%%%%%%%

Moreover, if the function $f$ is such that $f (x) = x f( x^{-1})$ and $f(1) = 1$, the correspondence between such functions and monotone metrics is one-to-one.

Making use of the projection map introduced in \eqref{eqn:projection_map_2} we can pull-back the monotone metrics in \eqref{eqn:Petz_characterization} to the unfolding space. In fact, we can define a twice covariant tensor $	G^f_\mathcal{M}$ on $\mathcal{M(H)}$ via
%%%%%%%%%%%%%%%%%%%%%%%%%%%%%%%%%%%%%%%%%%
\begin{equation}
G^f_\mathcal{M} := \pi^* G^f,
\end{equation}
%%%%%%%%%%%%%%%%%%%%%%%%%%%%%%%%%%%%%%%%%%
this means that the action of $	G^f_\mathcal{M}$ on two tangent vectors can be written as
%%%%%%%%%%%%%%%%%%%%%%%%%%%%%%%%%%%%%%%%%%
\begin{equation}
%%%%%%%%%%%%%%%%%%%%%%%%%%%%%%%%%%%%%%%%%%
\begin{split}
\left( G^f_\mathcal{M} \right)_{( U, \mathbf{p})} \left( (i H, v), (i K, u) \right) :&= \left( \pi^* G^f \right)_{( U, \mathbf{p})}\left( (i H, v), (i K, u) \right)\\
& = (G^f)_{\pi ( U, \mathbf{p})} \left( T_{( U, \mathbf{p})} \pi (i H, v),T_{( U, \mathbf{p})} \pi (i K, u) \right),
\end{split}
%%%%%%%%%%%%%%%%%%%%%%%%%%%%%%%%%%%%%%%%%%
\end{equation}
%%%%%%%%%%%%%%%%%%%%%%%%%%%%%%%%%%%%%%%%%%
where $T_{( U, \mathbf{p})} \pi$ denotes the tangent map of $\pi$ at the point $ ( U, \mathbf{p})$, $H$ and $K$ are self-adjoint operators on $\hh$ and $u,v \in T_{\mathbf{p}}\Delta^{\mathrm{o}}_{n-1}$.

%%%%%%%%%%%%%%%%%%%%%%%%%%%%%%%%%%%%%%%%%%
\begin{remark}
	A crucial thing to notice is that the tensor $G^f_\mathcal{M}$ is not a metric tensor, since it has a non-trivial kernel. To see this, consider a curve $m(t) = ( U(t), \mathbf{p})$ on $\mathcal{M} ( \hh)$ such that $\mathbf{p}$ does not depend on $t$, while  $U(t)$ is in the isotropy group of $\pi( m(0))$ for all $t$, then holds
	%%%%%%%%%%%%%%%%%%%%%%%%%%%%%%%%%%%%%%%%%%
	\begin{equation}
	\pi( m(t)) = \pi( m(0)) := \tilde{ \rho} \quad \forall t,
	\end{equation}
	%%%%%%%%%%%%%%%%%%%%%%%%%%%%%%%%%%%%%%%%%%
	implying that for any vector $V_\pi$ tangent to such curve we have
	%%%%%%%%%%%%%%%%%%%%%%%%%%%%%%%%%%%%%%%%%%
	\begin{equation}
	T_{ m(0)} \pi ( V_\pi) = \mathbf{0},
	\end{equation}
	%%%%%%%%%%%%%%%%%%%%%%%%%%%%%%%%%%%%%%%%%%
	where $\mathbf{0}$ is the null vector in $T_{\tilde{ \rho}} \stsph$. This implies
	%%%%%%%%%%%%%%%%%%%%%%%%%%%%%%%%%%%%%%%%%%
	\begin{equation}
	\left( G^f_\mathcal{M} \right)_{m(0)} \left( V_\pi, V_\pi \right) = (G^f)_{\tilde{ \rho}} \left( \mathbf{0}, \mathbf{0} \right) = 0.
	\end{equation}
	%%%%%%%%%%%%%%%%%%%%%%%%%%%%%%%%%%%%%%%%%%
\end{remark}
%%%%%%%%%%%%%%%%%%%%%%%%%%%%%%%%%%%%%%%%%%

In \cite{C-DC-DN-V-2022}, the authors find that the tensor $G^f_{\mathcal{M}}$ can be written as 
%%%%%%%%%%%%%%%%%%%%%%%%%%%%%%%%%%%%%%%%%%
\begin{equation}
\left( G^f_\mathcal{M} \right)_{( U, \mathbf{p})} \left( ( iH, v), ( iK, u) \right) = G^f_{\mathcal{U}} ( iH, iK) + \pi_D^* \left( G^{n-1}_{FR} ( v, u) \right).
\end{equation}
%%%%%%%%%%%%%%%%%%%%%%%%%%%%%%%%%%%%%%%%%%
Let us comment on this result in some detail: here $G_{FR}^{n-1}$ is the Fisher-Rao metric tensor defined on $\Delta^{\mathrm{o}}_{n-1}$, $\pi_D$ is the dequantization map defined as in \eqref{eqn:definition_dequantization_map}, so that the last term of this expression can be considered as some sort of classical term of the metric $G^f_\mathcal{M}$. Moreover, the term containing $G^{n-1}_{FR}$ does not depend on the choice of the operator monotone function $f$, meaning that this is a feature of all metrics that can be obtained via the unfolding procedure from the one introduced by Petz. The term $G^f_{\mathcal{U}} ( iH, iK)$, on the other hand, depends on the choice of the operator monotone function $f$ and it cannot be seen as the pull-back of a tensor defined on $\Uh$, since it depends also on $\mathbf{p}$. Nonetheless, it can be shown that this term is equal to zero whenever $H$ and $K$ commute with $\pi( U, \mathbf{p})$.

This shows how Classical Information Geometry can in a sense always be considered as contained in Quantum Information Geometry. More specifically, whenever one considers vectors in $T_{( U, \mathbf{p})} \mathcal{M}$ commuting with $\pi ( U, \mathbf{p})$ one recovers exactly the only tensor that is relevant in Classical Information Geometry. 

As already recalled in the introduction, it is possible to regard the Fisher-Rao metric as a ``second-order expansion'' of Kullback-Leibler relative entropy. Now, adapting the results in \cite{L-R-1999} to our framework, we want to show that if we pull-back \emph{relative $g$-entropies} from the space of quantum states to the unfolding space and take some sort of ``second-order expansion'' of it, we recover equation \eqref{eqn:Petz_characterization}, i.e. the unfolded version of Petz's characterization of monotone metrics. 

\emph{Relative $g$-entropies} are defined as two-point functions on the space of quantum states,
%%%%%%%%%%%%%%%%%%%%%%%%%%%%%%%%%%%%%%%%%%
\begin{equation}
H_g: \stsph \times \stsph \ni ( \rho, \sigma) \mapsto S_g ( \rho, \sigma) = \Tr \left( \sqrt{\rho} g ( \mathbf{L}_{\sigma} \mathbf{R}_{\rho^{-1}} ) (\sqrt{\rho}) \right) \in \mathbb{R},
\end{equation}
%%%%%%%%%%%%%%%%%%%%%%%%%%%%%%%%%%%%%%%%%%
where $g$ is an operator convex function. It turns out that $H_g$ satisfies the monotonicity condition
%%%%%%%%%%%%%%%%%%%%%%%%%%%%%%%%%%%%%%%%%%
\begin{equation} \label{eqn:Les_Rus_entropies}
H^\hh_g ( \rho, \sigma) = H^{\mathcal{K}}_g ( \Phi( \rho), \Phi( \sigma))
\end{equation}
%%%%%%%%%%%%%%%%%%%%%%%%%%%%%%%%%%%%%%%%%%
for all Hilbert spaces $\hh$ and $\mathcal{K}$ and for all CPTP maps $\Phi:\bh \rightarrow \mathcal{B ( K)}$ \cite{C-I-R-R-S-Z-1993}
Let now $( U, \mathbf{p})$ and $( V, \mathbf{q})$ be two points in $\mathcal{M} ( \hh)$ such that
%%%%%%%%%%%%%%%%%%%%%%%%%%%%%%%%%%%%%%%%%%
\begin{equation}
%%%%%%%%%%%%%%%%%%%%%%%%%%%%%%%%%%%%%%%%%%
\begin{split}
\rho = \pi( U, \mathbf{p}), \\
\sigma = \pi( V, \mathbf{q}),
\end{split}
%%%%%%%%%%%%%%%%%%%%%%%%%%%%%%%%%%%%%%%%%%
\end{equation}
%%%%%%%%%%%%%%%%%%%%%%%%%%%%%%%%%%%%%%%%%%

In order to pull-back two-point functions we need to define a new projection map that extends the projection map $\pi$ to the cartesian product of $\mathcal{M(H)}$ with itself, i.e. 
%%%%%%%%%%%%%%%%%%%%%%%%%%%%%%%%%%%%%%%%%%
\begin{equation}
\Pi: \mathcal{M (H)} \times \mathcal{M (H)} \ni ( ( U, \mathbf{p}), ( V, \mathbf{q}) ) \mapsto ( \pi ( U, \mathbf{p}), \pi ( V, \mathbf{q})) \in \stsph \times \stsph.
\end{equation}
%%%%%%%%%%%%%%%%%%%%%%%%%%%%%%%%%%%%%%%%%%
In this way $\Pi^* S^\hh_g$ is a two-point function on $\mathcal{M (H)}$, a short computation shows that
%%%%%%%%%%%%%%%%%%%%%%%%%%%%%%%%%%%%%%%%%%
\begin{equation}
\Pi^* \left( S^\hh_g \right) \left( ( U, \mathbf{p}), ( V, \mathbf{q}) \right) = \sum_{j,k} g \left( \frac{q^j}{p_k} \right)  \bra{e_k} U^\dagger V \ket{e_j} \bra{e_j} V^\dagger U \ket{e_k}.
\end{equation}
%%%%%%%%%%%%%%%%%%%%%%%%%%%%%%%%%%%%%%%%%%
Clearly the second derivatives will now be carried out using the tools of differential geometry and in particular, for the part containing the unitary operators, making use of the differential calculus on Lie groups. Here we will not retrace the computations and refer the reader to \cite{C-DC-DN-V-2022} for details. This gives rise to the expression
%%%%%%%%%%%%%%%%%%%%%%%%%%%%%%%%%%%%%%%%%%
\begin{equation} \label{eqn:computation_Les_Rus}
%%%%%%%%%%%%%%%%%%%%%%%%%%%%%%%%%%%%%%%%%%
\begin{split}
G^{ \mathcal{M} }_{ g }(X,Y)   & = g''(1)\,\pi_r^{*} G_{FR}^{n-1} (X,Y)   \\
&+ g\left( 1\right)\sum_{j=1}^{k}  p^j \bra{e_j} U^\dagger \mathrm{d}U (X) U^{\dagger}\mathrm{d}U(Y) + U^\dagger \mathrm{d}U (Y) \, U^\dagger \mathrm{d}U(X)  \ket{e_j}  \\
& - 2g\left(1  \right) \sum_{j=1}^{n} p^j \,  \bra{e_j} U^\dagger \mathrm{d}U(X)  \ket{e_j} \bra{e_j} U^\dagger \mathrm{d}U (Y) \ket{e_j}  \\
& -2 \sum_{k>j} \left(g\left( \frac{p^j}{p^k}  \right) p^k + g\left( \frac{p^k}{p^j}  \right) p^j \right)  \,\,\Re\left(\bra{e_k} U^\dagger \mathrm{d}U (X) \ket{e_j} \bra{e_j} U^\dagger \mathrm{d}U(Y)  \ket{e_k}\right)  .
\end{split}
%%%%%%%%%%%%%%%%%%%%%%%%%%%%%%%%%%%%%%%%%%
\end{equation}
%%%%%%%%%%%%%%%%%%%%%%%%%%%%%%%%%%%%%%%%%%
Where $X$ and $Y$ are in $T_{ ( U, \mathbf{p})} \mathcal{M (H)}$ and $U^\dagger \mathrm{d} U$ is the Maurer-Cartan one-form \cite{Cartan-1904} on $\Uh$.

It can be shown that this expression coincides with \eqref{eqn:Petz_characterization} iff $g(1) = 0$ and 
%%%%%%%%%%%%%%%%%%%%%%%%%%%%%%%%%%%%%%%%%%
\begin{equation} \label{eqn:relation_f_and_g}
f(x) = \frac{ ( 1 - x )^2 }{ g( x ) + x g ( x^{-1} ) }.
\end{equation}
%%%%%%%%%%%%%%%%%%%%%%%%%%%%%%%%%%%%%%%%%%
holds. Using the last expression it is also easily seen that $g''(1) = f(1) = 1$. Equation \eqref{eqn:relation_f_and_g} provides a relation between the operator convex function $g$ in \eqref{eqn:Les_Rus_entropies} and the operator monotone function $f$ in \eqref{eqn:Petz_characterization}.

This result allows to simplify significantly the expression in \eqref{eqn:computation_Les_Rus},
%%%%%%%%%%%%%%%%%%%%%%%%%%%%%%%%%%%%%%%%%%
\begin{equation}
%%%%%%%%%%%%%%%%%%%%%%%%%%%%%%%%%%%%%%%%%%
\begin{split}
G^{ \mathcal{M} }_{ g }(X,Y) & = \pi_D^{*} G_{FR}^{n-1} (X,Y) \\ 
& -2 \sum_{k>j} \left(g\left( \frac{p^j}{p^k}  \right) p^k + g\left( \frac{p^k}{p^j}  \right) p^j \right)  \,\,\Re\left(\bra{e_k} U^\dagger \mathrm{d}U (X) \ket{e_j} \bra{e_j} U^\dagger \mathrm{d}U(Y)  \ket{e_k}\right),
\end{split}
%%%%%%%%%%%%%%%%%%%%%%%%%%%%%%%%%%%%%%%%%%
\end{equation}
%%%%%%%%%%%%%%%%%%%%%%%%%%%%%%%%%%%%%%%%%%
plugging \eqref{eqn:relation_f_and_g} in the last expression we get
%%%%%%%%%%%%%%%%%%%%%%%%%%%%%%%%%%%%%%%%%%
\begin{equation}
%%%%%%%%%%%%%%%%%%%%%%%%%%%%%%%%%%%%%%%%%%
\begin{split}
G^{ \mathcal{M} }_{ g } = \pi_D^{*} G_{FR}^{n-1} -2 \sum_{k>j} \left(p_k - p_j \right)^2  \left( p_k \, f\left( \frac{p^j}{p^k}  \right)\right)^{-1}  \,\,\Re\left(\bra{e_k} U^\dagger \mathrm{d}U \ket{e_j} \otimes \bra{e_j} U^\dagger \mathrm{d}U  \ket{e_k}\right).
\end{split}
%%%%%%%%%%%%%%%%%%%%%%%%%%%%%%%%%%%%%%%%%%
\end{equation}
%%%%%%%%%%%%%%%%%%%%%%%%%%%%%%%%%%%%%%%%%%
This expression is in complete agreement with \eqref{eqn:Petz_characterization}, for a complete proof of this statement see \cite{C-DC-DN-V-2022}. As anticipated, we can see that the expression exhibits a splitting in two terms, one containing Fisher-Rao metric tensor and the other is a term that doesn't appear in the classical case and it is a two-contravariant tensor that depends on the choice of the operator monotone function $f$ and it is not the pull-back of some tensor defined on the unitary group $\Uh$. This can be seen from the fact that also the second terms depends on the values of the components of $\mathbf{p}$.
%%%%%%%%%%%%%%%%%%%%%%%%%%%%%%%%%%%%%%%%%%
%%%%%%%%%%%%%%%%%%%%%%%%%%%%%%%%%%%%%%%%%%
%%%%%%%%%%%%%%%%%%%%%%%%%%%%%%%%%%%%%%%%%%
%%%%%%%%%%%%%%%%%%%%%%%%%%%%%%%%%%%%%%%%%%
%%%%%%%%%%%%%%%%%%%%%%%%%%%%%%%%%%%%%%%%%%
\section{Conclusions}
%%%%%%%%%%%%%%%%%%%%%%%%%%%%%%%%%%%%%%%%%%
We showed how the shift from the state of quantum states $\mathcal{S(H)}$ to the unfolding space $\mathcal{M(H)}$ makes it strikingly easy to recognize at first glance the classical structures underlying the quantum world. Stated differently, some sort of dequantization process becomes possible by simply forgetting about the first argument in the couple $(U, \boldsymbol{p})$. In fact, we showed the expression that the metric tensors characterized by Petz assume in this framework, and we saw that in the unfolding space the metric tensors split in two terms. One contains information about the classical aspects and, as one would expect, is exactly Fisher-Rao metric. The other one is a term that doesn't appear in the classical case and it is a two-contravariant tensor depending on the choice of the operator function $f$ and written containing the contribution given by the action of the unitary group.

Finally we recalled the procedure of obtaining a monotone quantum metric from a relative $g$-entropy in this unfolded perspective. Again the results are in agreement both with the original result in \cite{L-R-1999} and with the unfolded version of Petz's characterization. With this approach we also saw how the ``non-classical term' 'is a two-contravariant tensor written in terms of the Maurer-Cartan one form on $\Uh$. This shows that, apart from the conceptual idea behind the unfolding, this can also be exploited for practical uses. In fact, this allows to set the whole discussion on a manifold which is the product of two manifolds whose geometry is well-studied. In particular, ideas from the differential calculus on Lie groups can be applied to perform calculations on $\Uh$.

It is also true that one can obtain dually-related connection from third order derivative of relative $g$-entropies. A possible next step would be to compute the family of dually-related connections determined by the relative g-entropies in this unfolded perspective. What one would expect is that also the skewness tensor describing this dual structure will split into a purely classical part and a quantum part given in terms of elements of the unitary group.

Another possible follow-up for this work would be to try to apply the unfolding procedure to the infinite-dimensional case. In this case the interior of the simplex will be replaced by the subset of the Banach space $l^1(\mathbb{R})$ given by strictly positive sequences adding to one. It is clear that such a generalization brings all the technicalities related to infinite-dimensional analysis and geometry, but at the same time it is immediate to see that such a generalization would allow to apply this procedure to relevant quantum mechanical problem such as the harmonic oscillator or the hydrogen atom.

\addcontentsline{toc}{section}{References}
\bibliographystyle{plain}
\bibliography{scientific_bibliography}

\end{document}